\documentclass[aps,pre,twocolumn,groupedaddress,showpacs]{revtex4-1}

\usepackage[colorlinks=true,urlcolor=black,linkcolor=black,citecolor=black,menucolor=black]{hyperref}
\usepackage{amsfonts,amssymb,amsbsy,amsmath}
\usepackage[pdftex]{graphicx}
\usepackage{MnSymbol}
\graphicspath{{./figures/}}

\begin{document}

\title{Effects of temporal correlations on cascades: Threshold models on temporal networks}

\author{Ville-Pekka Backlund}
\email[]{ville-pekka.backlund@aalto.fi}
\affiliation{BECS, School of Science, Aalto University, P.O. Box 12200, FI-00076 AALTO, Finland}

\author{Jari Saram\"{a}ki}
\email[]{jari.saramaki@aalto.fi}
\affiliation{BECS, School of Science, Aalto University, P.O. Box 12200, FI-00076 AALTO, Finland}

\author{Raj Kumar Pan}
\email[]{rajkumar.pan@aalto.fi}
\affiliation{BECS, School of Science, Aalto University, P.O. Box 12200, FI-00076 AALTO, Finland}

\date{\today}

\begin{abstract}
A person's decision to adopt an idea or product is often driven by the decisions of peers, mediated through a network of social ties. A common way of modeling  adoption dynamics is to use threshold models, where a node may become an adopter given a high enough rate of contacts with adopted neighbors. We study the dynamics of threshold models that take both the network topology and the timings of contacts into account, using  empirical contact sequences as substrates. The models are designed such that adoption is driven by the number of contacts with different adopted neighbors within a chosen time. We find that while some networks support cascades leading to network-level adoption, some do not: the propagation of adoption depends on several factors from the frequency of contacts to  burstiness and timing correlations of contact sequences. More specifically, burstiness is seen to suppress cascades sizes when compared to randomised contact timings, while timing correlations between contacts on adjacent links facilitate cascades.

%
%
\end{abstract}
\pacs{89.75.-k,02.50.Ey,89.75.Hc,05.45.Tp}

\maketitle

\section{\label{sec:intro} Introduction}

We live in a highly complex and interdependent world where every day we make a number of decisions.
Although individuals are free to make their decisions independently, it is a common human tendency to match one's attitude, beliefs and behavior to those of his or her acquaintances~\cite{cialdini_social_2004,salganik_musicmarket,bakshy_fb,castellano1}. Because of this, the ways how our social ties are structured and interlinked become important for understanding processes of social influence, especially system-level phenomena like the emergence of fads or adoption of ideas or behaviors. Such processes can then be approached quantitatively with models that define how individuals react to their neighbors' behavior, mediated through the links of some chosen type of social network where the individuals are embedded.

The most elementary models of social influence allow only two states for each individual (that is, network node): e.g., the node has either adopted an idea  or it has not. Then, the rules obeyed by each node dictate how they respond to their neighbors' choices. At the very simplest, social influence can then be modeled so that for a non-adopter, a single adopted neighbor can infect the node with adoption, leading to the standard epidemiological models such as SI (Susceptible-Infectious) or SIR (Susceptible-Infectious-Recovered)~\cite{anderson_sirbook} that adds a third possible state (Recovered). Even though these models are useful for studying the basic adoption process, they lack certain crucial features needed in the context of social contagion. As an example, adopting a product or behaviour is more likely if information is received from multiple adopted neighbors~\cite{centola_spread, ugander2012structural}. Thus, a more realistic choice of the node model is to use a \emph{threshold} model~\cite{granovetter_threshold,watts_casc}, where a node becomes an adopter only if a chosen fraction of its neighbors have already adopted, that is, where the level of peer pressure matters. The introduction of adopted individuals to a network of non-adopters may then eventually lead to a cascade where the majority of nodes become adopters; whether this happens or not depends on the model parameters and the underlying network structure. 

However, it has recently been realized that the above picture becomes more complete when an additional dimension is included: time. Social influence is transmitted through social ties, and in reality, they are not always active -- rather, there is a time series of contacts, from face-to-face conversations to emails. Incorporating the times of contacts into the network picture leads to the temporal networks framework~\cite{holmesaramaki_rev_temporal}, where the nodes  are connected only by temporary events of short duration at specific times. Studies of time series of contacts, based on electronic communication records, have shown that  contact timings display strong heterogeneity and burstiness, and that this burstiness has strong effects on dynamics taking place on temporal networks. In particular, burstiness often slows down spreading processes such as SI or SIR~\cite{karsai_smallbutslow,kivela_multiscaleanalysis,miritello_dynamical_2011}.

It is then natural to ask how cascades occur in  threshold models if node behavior depends on the exact timings of its interactions. As there is now an additional degree of freedom -- time -- to be incorporated into the models, there are several possible approaches. The original threshold model by Watts~\cite{watts_casc} is purely topological, in the sense that the dynamic process of adoption takes place on a static network topology. In the other limit, the threshold models proposed by Karimi and Holme~\cite{karimi_threshold} and Takaguchi \emph{et al.}~\cite{takaguchi_bursty} can be seen as purely temporal: what drives the adoption of a node is the number of recent contacts from adopted individuals, such that multiple contacts from the same adopted individual have the same effect as the same number of contacts from multiple adopted sources. Thus the role of the underlying network topology is diminished. However, one could argue that in a temporal threshold model that is explicitly designed to mimic cascades in social networks, the underlying network topology should be accounted for, as the effects of social reinforcement should be stronger when the signal arrives from multiple distinct acquaintances (for empirical results, see~\cite{centola_spread}). In other words, adoption should be more likely if several friends have already adopted.

To this end, in this paper we propose two threshold models -- stochastic and deterministic -- that could be categorized as temporal-topological. In these models, adoption is driven by the number of contacts with a node's different network neighbors within a chosen time window representing node memory: a high number of contacts from different adopters makes adoption more likely. In this respect, the models are similar to the general model of contagion of Ref.~~\cite{doddswatts_contagion}. However, our models take the actual timings of interactions into account and are therefore affected by any heterogeneities of the activity patterns of nodes, such as burstiness. Note that the thresholds are defined in terms of numbers of neighbours of nodes during the whole available timeline, \emph{i.e.} their static degrees. The rationale behind this assumption is the separation of time scales: we assume that the time scale of changes \emph{of} the network (creation and deletion of nodes and links) is much slower than the time scale of the adoption process taking place \emph{on} the network. This choice also allows studying the effects of heterogeneities in contact timings in isolation.


Using empirical temporal network datasets, we then study with simulations the conditions leading to global cascades of adoption in these models. We found that for the stochastic model, long enough memory windows facilitated global cascades in only two out of four empirical networks. When a hard threshold is imposed on the fraction of adopted neighbors, global cascades appear only if this threshold is tuned so low  that for most nodes a single contact with a single adopted neighbor triggers adoption. This is because high-degree nodes block adoption, unlike for typical spreading processes (SI, SIR). To pinpoint the effects of different types of temporal correlations, we compare the results with those obtained with reference models that remove such correlations from the empirical data. We find that for our models, burstiness decreases adoption prevalence, whereas timing correlations between contacts on adjacent links clearly facilitate adoption.

\section{Methods}
\label{sec:methods}

\subsection{Basic Framework and Model Definition}
\label{sec:models}
The set of $\mathcal E$ events representing the interactions occurring between $N$ nodes within the time interval $[0,T]$ constitutes a temporal network. In this network, each  event is denoted by a quadruplet $e \equiv (u, v, t, \delta t)$, where the event connecting nodes $u$ and $v$ begins at $t$ and ends at $t+\delta t$. For instance, for call data, each event in the temporal network would correspond to a list of the caller, callee, starting time and duration of the call. The corresponding static, aggregated network is obtained by ignoring  temporal information and linking a pair of nodes if any event occurs between them in the observed period ($t\in [0,T]$). One can then view the events of the temporal networks as temporary activations of the links of the underlying static network: events define which links of the static network are active at any given point in time.

Our topological-temporal threshold models for node behavior, defined below, take into account both the timings of events and the aggregated network structure in terms of network neighborhoods. In the first model, there is an element of randomness in the adoption choices of nodes, whereas the second model is entirely deterministic. In both models, nodes have two possible states, susceptible and adopter, and initially all the nodes in the temporal network are set as susceptible except for one randomly chosen seed adopter node. When susceptible nodes become adopters, they remain so for the rest of the time.

\begin{figure}[b]
    \includegraphics[width=8cm]{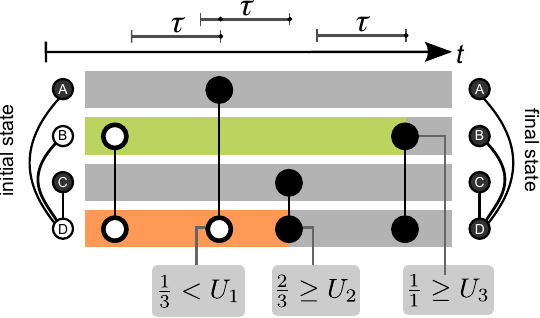}
    \caption{(color online) Schematic representation of the stochastic model of adoption dynamics. In the beginning of this example run, nodes A and C are adopters. Time runs from left to right, and each contact between a pair of nodes is denoted by a vertical line. The intervals $\tau$ illustrate the memory associated with each node. The lower panels show the adoption probabilities $\phi$ when a node is in contact with an adopter; $U_1$, $U_2$ and $U_3$ are random numbers associated with the event. For the event occurring between nodes A and D, only 1 out of 3 neighbors is an adopter and the ratio 1/3 is smaller than the random number $U_1$, and hence D does not adopt. Next, for the event between C and D, two of the neighbors of D who had contacts within the memory interval $\tau$ are adopters (A and C). Since the drawn random number  $U_2<2/3$, D adopts. Finally, since B has only one neighbor (D), it also becomes an adopter when in contact in D, regardless of the value of the random number $U_3$.}
    \label{fig:model}
\end{figure}

{\textbf{-Stochastic Threshold Model.}}
In the stochastic topological-temporal threshold model, the probability of a susceptible node becoming an adopter  depends on the number of contacts from different adopted nodes in a given time window: more adopted neighbors in contact means higher probability. Formally, when a node is in contact with an adopter at time $t$, the number of adopted neighbors that it has been in contact with during the interval $[t-\tau,\ t]$ is counted. For directed networks, only incoming contacts, that is, contacts initiated by the neighbors of the node are counted. The ratio of the number of contacted neighbors to the number of neighbors in the aggregated network, i.e., the degree $k_i$, determines the probability of adoption:
\begin{equation}
\phi(i,t|\tau):= \frac{1}{k_i} \sum_{j~\in~\nu_i} \chi (j,t') \quad \quad t'\in[t-\tau,\ t].
\label{eq:inf3}
\end{equation}
Here, $\nu_i$ the set of neighbors of node $i$ in the aggregated static network. The indicator function $\chi$ is 1 if node $j$ is an adopter and it has had at least one contact with node $i$ within the interval $[t-\tau,\ t]$, and 0 otherwise. The summation runs over all the neighbors of node $i$. The model comes with only one parameter -- the time window size, or memory length, $\tau$.

For a given node, the likelihood of adoption increases monotonically with the fraction of adopted neighbors, which is a reasonable assumption~\cite{centola_spread}.  Because the threshold is defined as fractional, the effect of the number of adopted neighbors depends on node degree.  Note that although the threshold rule does not directly count the number of contacts from the same adopted neighbor, these still contribute indirectly because the stochastic rule is activated whenever a contact occurs. Thus bursts of events affects the dynamics. For a schematic illustration of the model, please see Figure~\ref{fig:model}.

{\textbf{-Deterministic Threshold Model.}}
We also consider a deterministic version of the above model. In this model, a node becomes an adopter if the ratio $\phi(i,t|\tau) \geq f$, where $f$ is a predefined constant and common for all nodes. By varying the value of $f$ one can control the ease with which adoption progresses. Because the threshold $f$ is constant, the degree $k_i$ of a node plays a major role: if $k_i\leq1/f$, a single contact with an adopted node is sufficient for adoption. For nodes with $k_i>1/f$, larger number of contacts from separate adopters is required. In the deterministic model, burstiness of events on individual links has no effect unless the event trains of adjacent links are correlated: if a node does not adopt when in contact with an adopter, subsequent contacts with the same adopter are redundant and can not cause adoption unless there are also contacts with other adopters.

Note the general difference between the models: given long enough times (and applying periodic temporal boundary conditions), the stochastic model would lead to a complete system wide adoption (similarly to the SI-model, albeit much more slowly). However, in the deterministic case, even though we can tune the effective number of nodes that only require a single contact with an adopter for adoption, the diffusion dynamics may be blocked entirely by a small number of nodes where the condition for adoption is not fulfilled.

\subsection{Datasets}
\label{sec:data}
In this study, we simulate the above threshold models on four different empirical temporal networks that are related to different types of human communication patterns. These datasets are:

\textit{Call:}
The call data contain mobile phone call records of an European carrier, with $\sim 759$ million time-stamped voice call records over a period of 180 days. The  durations of the calls are ignored in this study.

\textit{SMS:} The text message data contains SMS records from the same European carrier and the same time period. There are $\sim 243$ million time-stamped records. We exclude data for Christmas and New Year's Eve since the behavior (in terms of the number of messages) on these two days is radically different from other days.

\textit{Email:}
The email dataset~\cite{eckmann_email} consist of logs of a university's mail server for a period of 83 days. Only inter-institute emails are considered and certain mass mailers are discarded.  

\textit{Conference:}
The conference dataset is a collection of ongoing face-to-face conversations of 113 participants of the ACM Hypertext 2009 conference for 2.5 days~\cite{isella_whatsinthecroud,sociopatterns}. The participants were tracked with the help of radio badges that monitored their proximity to other attendees. The data contains timestamps and IDs of pairs of participants who are having a conversation within time windows of 20 seconds. 

\begin{table}[b]
\caption{Statistical properties of the networks. From left to right: number of nodes $N$, number of events $E$, number of links $m$, average degree $\langle k \rangle$, average clustering coefficient $\langle c \rangle$, length of the observation period $T$, time resolution of the data $\Delta t$ and the directionality $\rightarrow$ or $\leftrightarrow$ of the links. $\langle k \rangle$ and $\langle c \rangle$ have been calculated for the aggregated networks by considering links as undirected.}
\centerline{
\begin{tabular}{c | c c c c c c c c}
		& $N$ & 		$E$ 				& $m$	& $\langle k \rangle$ & $\langle c \rangle$ & $T$ & $\Delta t$ & link \\ \hline 
Call		& 5.8$\times 10^6$	&620$\times 10^6$		&15$\times 10^6$		&5.0	&0.23	&180 d	&1 s & $\leftrightarrow$\\
SMS		& 3.1$\times 10^6$	&180$\times 10^6$		&5.7$\times 10^6$	&3.7	&0.10	&176 d	&1 s & $\rightarrow$\\
Email		& 2993			&2.0$\times 10^5$		&21736			&14.5	&0.21	&82 d	&1 s & $\rightarrow$\\
Conf.		& 113			&20818				&2196			&38.9	&0.54	&2.5 d	&20 s & $\leftrightarrow$\\
\hline
\end{tabular}}
\label{t:net_stats}
\end{table}
There are certain similarities and differences between these datasets. The first three represent electronic communication at distance, while in the conference data contacts require physical proximity. Another difference is that the call and conference conversation contacts require activity from both participants, e.g., a call needs to be picked up before it can be recorded, whereas in the SMS and email networks only the sender's activity matters. For similar reasons, we consider call and conference networks as undirected, whereas the SMS and email networks are directed. Further, in the call data one node can only participate in one call at any given point in time, whereas simultaneous contacts with multiple nodes are 
possible in the SMS, email and conference networks. 

In the call and SMS networks we only consider those links that have bidirectional events to ensure that non-social interactions, such as telemarketing are ignored. Further, we only use the nodes in the largest connected component (LCC) of the aggregated network. For the email network the directionality of links was ignored while calculating the LCC. The properties of all datasets and corresponding aggregated networks are presented in Table~\ref{t:net_stats}. The average degree and the average number of events per node are the highest for the conference network and lowest for the SMS network.

\subsection{Reference models}
\label{sec:shuffle}
In order to highlight the effects of different types of temporal correlations on the model dynamics, we employ the reference model approach~\cite{holme_reachability,karsai_smallbutslow,kivela_multiscaleanalysis}. In this approach, the original event sequences are randomized such that chosen temporal and topological characteristics are retained, while some temporal correlations are lost. Then, the outcomes of model runs using original and randomized reference networks are compared. 
We consider the following two null models:

\textbf{-Random time shuffle (RTS).}
The time stamps of all the events are randomly shuffled. This destroys all temporal correlations of events within links and of the nodes' activity patterns, e.g., burstiness, as well as correlations between the timings of events on adjacent links. The time variation of the global activity rate is preserved, such as the typical daily pattern. 

\textbf{-Random offset (RO).}
All the events occurring on a given link are shifted by a time offset $\Delta t \in [0,T]$, chosen at random for each link. Periodic temporal boundary conditions are used, i.e., the time of each of the event is mapped back to the period $[0,T]$ by using a $\mod~T$ operation on the shifted time. As event sequences of different links are shifted by different offset the correlations between events on adjacent links, i.e., {\it link-link correlations}  are destroyed while temporal inhomogeneities within individual links, such as burstiness, are retained. By extension, all larger sequences of subsequent events, such as temporal motifs~\cite{kovanen_motifs_1,kovanen_motifs_2} are also destroyed, as are the overall activation timelines of nodes.

\begin{figure}[t]
    \includegraphics[width=8cm]{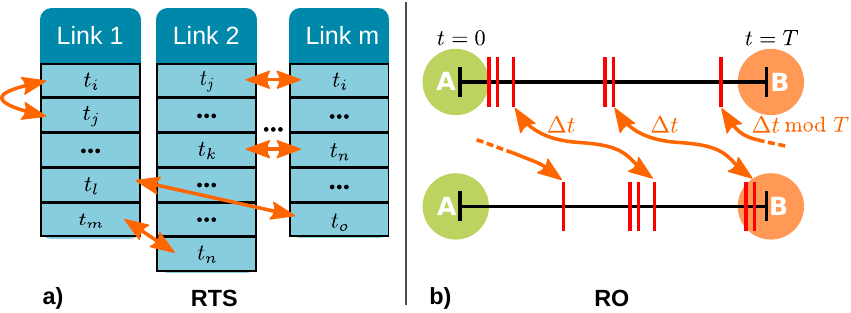}
    \caption{(color online) The event sequence shuffling procedures for the two reference models. (a) In random time shuffling (RTS), the time stamps of events are randomly swapped. (b) In random offset (RO), all events on a given link are shifted by a random time $\Delta t$. Periodic boundaries are used, i.e., if the resulting times fall outside the observed window [0,T]  a $\mod T$ operation maps them back to the period [0,T].}
    \label{fig:shuffles}
\end{figure}

For a schematic diagram of the shuffling procedures, see Figure~\ref{fig:shuffles}. Table~\ref{t:shuffles} summarizes the key effects of the null models on temporal correlations.

\begin{table}[t]
\caption{Network correlations that are preserved ($\checkmark$) or destroyed (x) by the two reference models.}
\centerline{
\begin{tabular}{| l || c | c | c |}
\hline
Correlations						& Orig.				& RTS				& RO 			\\ \hline
Network structure					& $\checkmark$		&	$\checkmark$	& $\checkmark$	\\
Daily patterns		      			& $\checkmark$		&	$\checkmark$	& x				\\
Single link event correlations		& $\checkmark$		&	x				& $\checkmark$	\\
Link-link event correlations	  	& $\checkmark$		&	x				& x				\\
\hline
\end{tabular}}
\label{t:shuffles}
\end{table}

\subsection{Simulations}
\label{sec:simulations}
We study adoption dynamics in the two threshold models by simulating them using the original event sequences and the ensemble of sequences from reference models. The initial adoption is started from a random node and random event within a time period that spans the first $15\%$ of the time period of the dataset. The model rules are then iterated until the end of the event sequence. The initial time span is chosen because a large enough fraction of nodes are active during this period (83\%, 71\%, 75\% and 71\% respectively for call, SMS, conference and email data) and there are enough events for the dynamics to advance. We calculate the global prevalence of adoption at the end of the event list $t=T$ as the ratio of the number of adopted nodes $I$ to the total number of active nodes $N_{\textrm{act}}$ that participated in at least one event during the run. $N_{\textrm{act}}$ practically equals network size, although some rare nodes may only be active before the random initial event. We obtain the mean prevalence by starting the adoption process from $10^3$ initial conditions for the larger call and SMS datasets, while for the smaller email and conference datasets we used $10^4$ initial conditions.


\section{\label{sec:results} Results}
\subsection{Stochastic model}

\begin{figure}[t]
    \includegraphics[width=8.6cm]{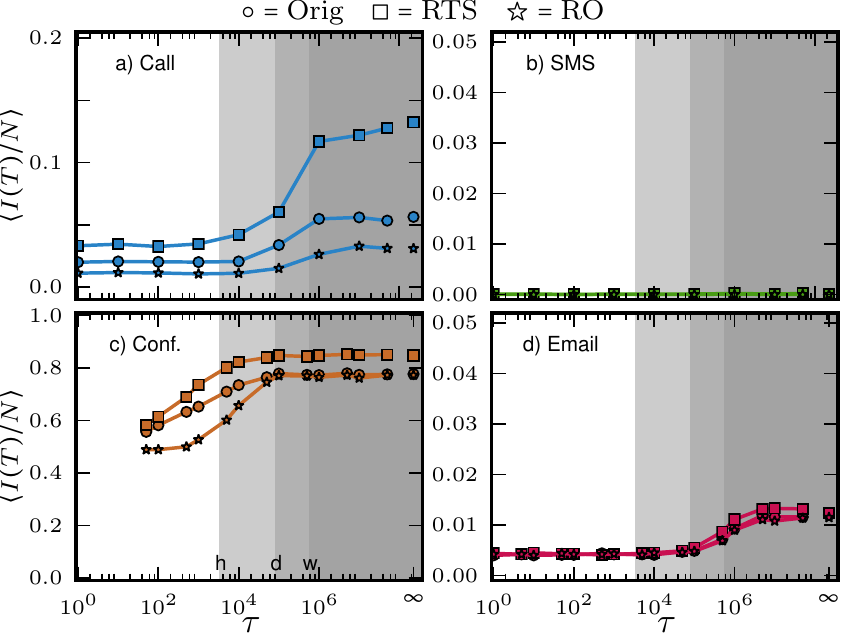}
    \caption{(color online) The average final fraction of adopters in the stochastic model as a function of memory length $\tau$  for (a) call, (b) SMS, (c) conference, and (d) email datasets. Curves are shown for the original event sequence ($\circ$) and the two null models: random time shuffling RTS ($\Box$) and random offset RO ($\medstar$). For all networks except SMS, there is a transition region where prevalence increases with memory length and then saturates. The prevalence is much lower for the SMS and email networks as compared to the call and conference networks. The RTS model facilitates adoption dynamics, whereas the RO model hinders adoption. Shades of grey represent memory lengths of one hour, one day and one week. Standard errors are smaller than symbol size. 
    }
    \label{fig:res_stochastic}
\end{figure}

\textbf{-Prevalence and memory length}. Figure~\ref{fig:res_stochastic} displays the outcomes of the stochastic model for all the four datasets, in terms of the average final fraction of adopters as a function of the memory parameter $\tau$, i.e. the length of the time window within which contacts from adopted individuals are counted. Overall, it is seen that with the shortest time windows, prevalence is low. When $\tau$ is increased, there is a transition regime where prevalence grows, followed by a plateau where increasing $\tau$ has only small effects. The low-$\tau$ behavior is as one might reasonably expect: increasing memory length means that there are more chances for contacts with adopters. 
Note that the contrary is observed in the model of Ref.~\cite{karimi_threshold}, where the number of contacts with adopters is normalized by the total number of contacts within the time window, and because of this, longer memory results in lower adoption rates. It is also of interest to look at the position of the transition region, as it defines the time scale for adoption. In the call network, having a memory $\tau$ less than one hour does not facilitate adoption, whereas increasing the memory to more than one week  becomes redundant. In the conference network, $\tau$ longer than one day becomes redundant. 
 
It is also clear that there are major differences between the datasets: for high enough values of $\tau$, the conference network has the highest prevalence, followed by the call network, whereas the fraction of adopters in the SMS and email networks always remains very low. We have tested that the fraction of adopters in the SMS network is low even when the directionality of the events is ignored. This indicates that the reason for the low fraction of adopters lies in the temporal and topological features of the networks rather than in the directionality of the links: the main reasons are the low link density of the network, the existence of large number of links with relatively low contact frequency, and the fact that conversations by text messages give rise to trains of repeated events only between same pairs of nodes. These factors lower the likelihood of having temporal paths via which the diffusion can spread~\cite{pan_temppathlengths}. In addition, it is unlikely that nodes whose links have only few events receive contacts from multiple sources, and because of this the probability of crossing the adoption threshold remains low.

\textbf{-Effects of temporal correlations.} 
In order to understand the effects of various kinds of temporal correlations on the adoption dynamics, we next focus on the effects of the reference models on the outcome. The average prevalence as a function of $\tau$ for the two reference models, random time shuffle (RTS) and random offset (RO), are also displayed in Figure~\ref{fig:res_stochastic}. 

Let us focus on the RTS reference model, and the two networks (conference, calls) where global cascades take place and the effects of reference models are best visible. Contrary to the absolute threshold model in Ref.~\cite{karimi_threshold} and the model in Ref.~\cite{takaguchi_bursty}, the RTS procedure that removes burstiness from the event sequences yields networks where the adoption rate is increased. 
This major difference to the purely temporal threshold models has to do with the effects of adoption from multiple neighbors. In the aforementioned models, bursts of contacts from a single neighbor can drive a node above the adoption threshold -- these bursts that facilitate spreading are destroyed by the RTS model. However, in our case, contacts from multiple adopted neighbors drive the adoption. As the RTS model spreads out bursty sequences of contacts more evenly in time, the likelihood of contacts with multiple neighbors in a given time window is increased. In addition, RTS model also increases the probability of having a time-respecting path between any two nodes at any given point of time, hence facilitating  adoption dynamics especially in the short memory region where multiple contacts are unlikely even with the reference model.

This picture becomes more complete with the results of the RO model. Compared to the Poissonian event sequences of the RTS model, burstiness of contacts in the original event sequences clusters them in time and decreases the likelihood of contacts with multiple neighbors within short time windows. However, if there are positive timing correlations between the bursty contact patterns of the links of a node, they should facilitate adoption. This is exactly what is seen in the prevalence curves from the RO model (Figure~\ref{fig:res_stochastic}) that randomly shifts the bursty event sequences on links, destroying timing correlations between adjacent links while retaining burstiness on individual links. For both the call and conference networks, applying the random offset results in a decrease of adoption prevalence. Hence, one can conclude that in the original event sequences for the call and conference networks, there are correlations between contact timings of adjacent links that facilitate adoption. At the same time, for the SMS and email networks, global cascades do not take place even for any of the reference models.

\begin{figure}[t]
    \includegraphics[width=8.6cm]{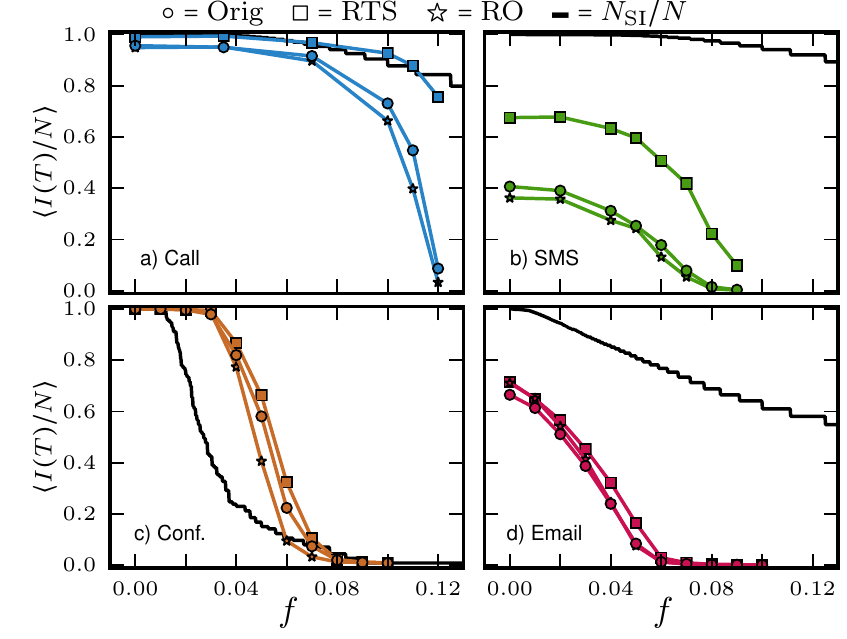}
    \caption{(color online) The average final fraction of adopters in the deterministic model as a function of fraction $f$ with $\tau=\infty$ for the (a) call, (b) SMS, (c) conference and (d) email datasets. The adoption process is shown for the original event sequence ($\circ$) and the two null models: (i) random time shuffle RTS ($\Box$) and (ii) random offset RO ($\medstar$). The black solid line represents the fraction of nodes in a network that obey pure SI-dynamics at given $f$ value. Except for the conference data, majority of the nodes must obey SI dynamics in order for a global diffusion to happen. Standard errors are smaller than symbol size.}
    \label{fig:res_threshold}
\end{figure}

\begin{figure}[t]
    \includegraphics[width=8.6cm]{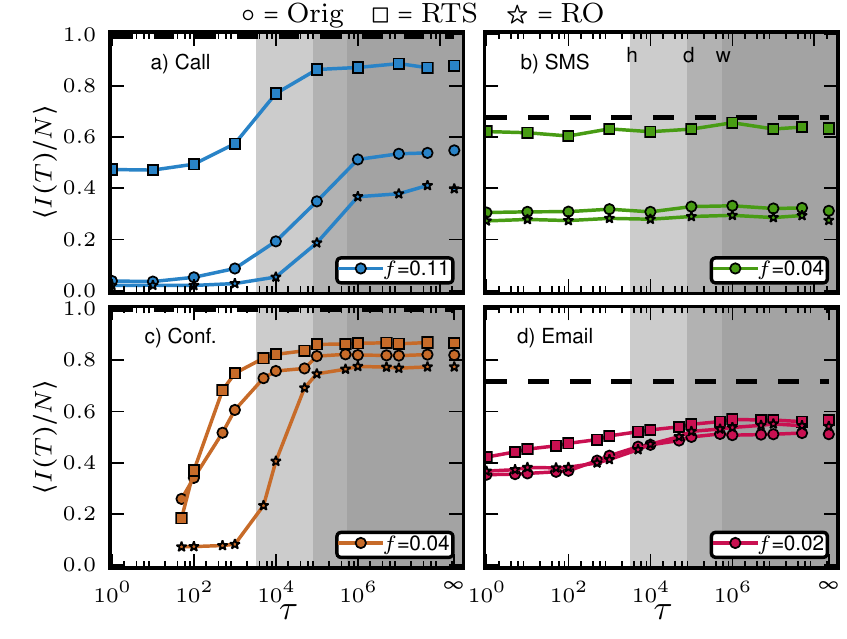}
    \caption{(color online) The average final fraction of adopters in the deterministic model as a function of memory $\tau$ for the (a) call, (b) SMS, (c) conference and (d) email datasets. Again, the adoption process is shown for the original event sequence ($\circ$) and the two null models: (i) random time shuffle RTS ($\Box$) and (ii) random offset RO ($\medstar$). The fraction $f$ is now chosen so that the curves have a transition as $\tau$ changes. The difference in the fraction of adopters in the RTS and RO null model is more evident in this case. As the temporal path do not exist between all the nodes, the maximum prevalence is also shown for all the datasets for RTS null model. Standard errors are smaller than symbol size. 
    }
    \label{fig:res_fractional}
\end{figure}

\subsection{Deterministic model}
\textbf{-Effects of $f$.}
Next, we consider the deterministic version of the threshold model, where adoption happens if and only if a fraction $f$ of the node's network neighbors are adopters and in contact with the node within the time period $\tau$. Hence, we can control the ease of adoption propagation by changing the threshold fraction $f$. When $f$ is very low, $f\sim1/k_\mathrm{{max}}$, where $k_\mathrm{{max}}$ denotes the maximum degree of the network, all events between adopted and susceptible nodes lead to an adoption. In this case, the model's dynamics are similar to the SI model. For any value of $f$, contact with a single adopter is sufficient for triggering adoption of a node if the node's degree $k\leq 1/f$. On the other hand, for large values of the threshold $f$, propagation of adoption would require that almost the whole neighborhood of a node are adopters and in contact with the node within the memory period $\tau$. This means adoption typically cannot propagate for large values of $f$. 

Fig.~\ref{fig:res_threshold} displays the fraction of adopters in the deterministic model as a function of $f$, together with the fraction of nodes with $k\leq 1/f$ that obey SI dynamics, $N_\mathrm{SI}/N$. In these simulations, the memory is fixed to $\tau=\infty$, so that the effect of $f$ can be determined in isolation.
 In all cases, the values of $f$ need to be small in order for the adoption to take off. For instance, in the call network adoption does not propagate at all if  $f\geq0.12$. This is noteworthy, since at this threshold fraction more than 80\% of the nodes follow SI dynamics and contact with a single adopter is sufficient for adoption. Only nodes with $k>8$ obey the true threshold mechanism. Because the majority of nodes follow SI dynamics and because pure SI dynamics would always infect the whole network, this indicates that high-degree nodes typically block the propagation of adoption, as they are unlikely to interact with enough adopters.
 For the SMS network, the adoption propagates only when $f\leq0.08$, with more than 95\% of the nodes following SI dynamics. In contrast, in the email network global adoption occurs at $f\leq0.04$, with about 80\% of nodes following SI dynamics. The differences imply that interactions with multiple neighbors are  more common in the email network as compared to the SMS network. Finally, for the conference network, global adoption arises even when most of the nodes follow pure threshold dynamics. This is because of the high frequency of interaction between nodes.

\textbf{-Effects of memory length $\tau$.}
Next, we study the behavior of the system as a function of the memory $\tau$. The results are shown in Figure~\ref{fig:res_fractional}, where the fraction $f$ for each dataset is chosen so that the adoption can propagate, given high enough $\tau$. Unlike with the stochastic model, even for the SMS and email networks, a substantial fraction of nodes become adopters in the end. As discussed above, this occurs at low thresholds, and most nodes obey SI dynamics, \emph{i.e.} a single contact with an adopted neighbor triggers adoption. This is also reflected in the observation that adoption prevalence in the SMS and email networks is barely affected by the memory $\tau$. For the call and conference networks the outcomes are rather similar to the stochastic model. The only difference is that the transition region where prevalence increases is associated with lower values of $\tau$. In the call network, this happens with $\tau\sim 1$ day, and for the conference network, $\tau\sim 1$ hour. For the stochastic model, the corresponding $\tau$ values are $\tau \sim 1$ week for calls and $\tau\sim 1$ day for the conference.

%

\textbf{-Effects of temporal correlations.}
The effects of the reference models for the deterministic model are consistent with the outcomes in the stochastic case. As before, for all $f$ the fraction of adopters increases for the RTS model and decreases for the RO model in all datasets. However, the deterministic case reveals larger differences between the reference models especially in the low $\tau$ region. This is because of the altered temporal path structure and its effects on the SI dynamics that most nodes obey. 

\begin{figure}[b]
    \includegraphics[width=8cm]{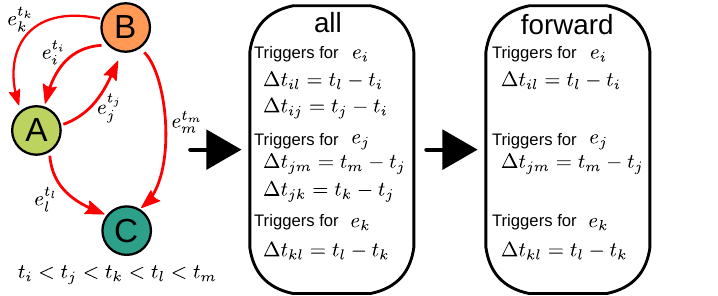}
    \caption{(color online) Calculation of the density of preceding events, for directed events. The red arrows represent the events; event times are indicated next to the arrows. All outgoing events from a node act as triggers for calculating time differences to preceding incoming events. The set of time differences for all nodes and all trigger events is then used to construct a probability density function.}
    
    \label{fig:act_trig}
\end{figure}

\subsection{Measuring temporal correlations}
\textbf{-Density of preceding events.} The above results point out that there are correlations between the timings of events on adjacent links, and their effects are especially strong for the call network. The existence of such correlations is also known from earlier studies on \emph{e.g.} temporal motifs~\cite{kovanen_motifs_1,kovanen_motifs_2} and inter-event time distributions~\cite{karsai_smallbutslow}. In order to directly measure these correlations and to look for possible characteristic time scales, we compute the density of preceding events (or event-triggered correlation function~\cite{LauriThesis,Spikes}). The target is to study what happens just before nodes participate in events. For directed events, we define the density of preceding events as follows: for every outgoing event of a node, the time differences $\Delta t$ to all its earlier incoming events are calculated (see Fig.~\ref{fig:act_trig}). The probability density functions for the $\Delta t$ values for all nodes and events then represent the average rate of incoming events preceding an outgoing event, and any timing correlations should be visible in this PDF. Especially, if there is a characteristic time for the correlations of events on adjacent links, this should be visible as a peak in the PDF. For undirected events, this PDF reduces to the usual distribution of inter-event times. For both undirected and directed events, we repeat the procedure separately for two cases: i) where preceding events with all neighboring nodes are taken into account (case \emph{all}), and ii) for three-node chains of events only, where $\Delta t$'s are computed only for those preceding events that do not involve the other party of the trigger event (case \emph{forward}). Thus, the difference between the \emph{all} and \emph{forward} cases is that in the latter, the effect of event cycles of length two, \emph{i.e.} returned events, is removed. 

\begin{figure}[t]
    \includegraphics[width=8.6cm]{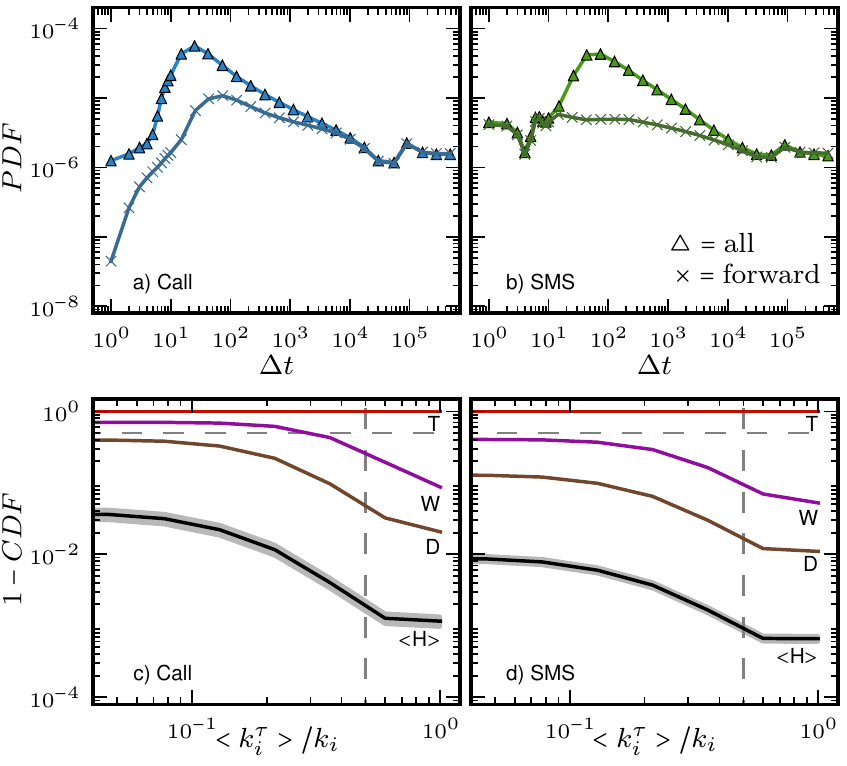}
    \caption{(color online) Density of preceding events (top row) and fraction of  accessible neighborhood in a given time window (bottom row) 
    for the call and SMS data.  Triangles and crosses indicate preceding event densities for all events and forward chains (events excluding the neighbor participating in the trigger event), respectively.  (a) For both cases, for the call network there is a peak in the distribution at around 25 seconds. (b) For the SMS data, the peak is only evident when all earlier events are considered -- as the SMS events are directed, this means that the peak in the preceding event density is mainly due to returned text messages. (c), (d) The complementary cumulative distribution of $\langle k_i^{\tau}\rangle/k_i$  for the call and SMS networks, respectively. $\langle k_i^{\tau}\rangle/k_i=0$ is not shown because the axis is logarithmic; the fraction of nodes with zero contacts is reflected in the starting point of the curves.  For any given time window the accessible neighborhood for the SMS network is much smaller than for the call network. The dashed lines indicate 50\% for both axes. The hourly CDF:s are first calculated independently for each hour of the day and then averaged;  the shaded areas represent the 75\% confidence interval for the mean of the 24 different CDF's. For longer memory intervals, the 75\% confidence interval for the mean is smaller than the line width. 
    }
\label{fig:res_eventcorr}
\end{figure}

 The preceding event densities for the call and SMS datasets are shown in Figure~\ref{fig:res_eventcorr}. There is a characteristic peak for event correlations on adjacent links: the PDF peaks at around $\sim25$~seconds for the call data, and at around $\sim45$~seconds for the SMS data when all events are considered. For forward event chains (events excluding the neighbor of the trigger event), the call network peak shifts to $\sim80$~seconds. In both cases, the time scales are fairly short, of the order of one minute, and thus event-event correlations have effects on the threshold model for a wide range of memory parameter values. For the directed text message network, there is no clear peak if only forward chains are counted. This is likely because text messages are typically part of a conversation, where two individuals repeatedly exchange multiple messages. These are redundant for adoption dynamics.
 
%

 \textbf{-Fraction of contacted neighbors and memory length}. In order to better understand the effects of the memory length on the two threshold models (stochastic, deterministic), it makes sense to directly measure how the amount of interactions with the neighborhood of a node depends on the memory window size. We count the number of nodes the focal node interacts with in a given window of size $\tau$ as $k_i[t,t+\tau]\equiv k_i^{\tau}$ and divide this number by the degree of the node in the aggregated static network, $k_i[0,T]\equiv k_i$. Averaging over nodes and windows yields the typical fraction of neighbors a node connects with within a given memory length $\tau$. Note that for the stochastic threshold model, the fraction $\langle k_i^{\tau}\rangle/k_i$ corresponds to the probability of adoption $\phi(i,\tau)$ when the whole neighborhood of a node has already adopted. In Figure~\ref{fig:res_eventcorr} we show the complementary cumulative distribution of $\langle k_i^{\tau}\rangle/k_i$ for the call and SMS networks for different memory lengths $\tau=$ (hour, day, week, the whole data period $T$). For the call network, it is clear why the memory window size has to be of the order of days for the adoption to propagate (Figs.~\ref{fig:res_stochastic},~\ref{fig:res_fractional}): for memory of one hour, the fraction of contacted neighbors is very low. For the SMS network, even for a window of one week, $\sim60$\% of the nodes have $\langle k_i^{\tau}\rangle/k_i$ less than 0.05, indicating rather infrequent communication.
 
\section{\label{sec:discussion} Discussion}
In this paper, we have introduced stochastic and deterministic versions of the topological-temporal threshold model of adoption, and studied their behavior using empirical temporal network datasets as substrates for the threshold dynamics. These models can be argued to reflect social diffusion processes better than some models proposed earlier~\cite{karimi_threshold,takaguchi_bursty}, since adoption probability directly depends on the number of contacted nodes, a mechanism observed in real-world experiments~\cite{centola_spread}. However, in addition to this, an equally important factor of motivation is that these models can be used as probes of temporal network structure: because of their design, our models are sensitive to timing correlations of the event trains on links.

For the stochastic threshold model, the call and conference networks allow global cascades for large enough memory windows (days for calls, hours for the conference network), whereas the SMS and email networks did not support such cascades because of the sparsity of interactions and lack of timing correlations between contacts on adjacent links. The characteristic window sizes are meaningful for the networks in question: in the face-to-face network recorded in a conference setting, people participate and switch between conversations over short time scales, whereas interactions via calls are less frequent. For the deterministic model, where adoption only takes place when the fraction of adopters in a node's neighbourhood exceeds a fixed threshold fraction, global adoption only took place when the threshold fractions were set to low enough values and most nodes effectively followed SI dynamics. This is consistent with the results of threshold dynamics on static networks, where low-degree vulnerable nodes drive the adoption dynamics~\cite{watts_casc}. One reason for this comes directly from model design: the higher the degree of a node, the larger the number of neighbors that first need to adopt and subsequently be in contact with the node within the memory window. Thus the effects of hubs are very different from ordinary (SI, SIR) spreading processes: while they may act as superspreaders once they have adopted, making them adopt is very difficult.

Randomly shuffling the times of all events in the empirical networks was seen to facilitate the adoption process and results in larger prevalence, similarly to earlier observations with the SI model~\cite{karsai_smallbutslow}. Thus, generally speaking, the burstiness of contacts that is ubiquitous in temporal networks hinders adoption because of increased waiting times on links and redundant repeated events. The time shuffling procedure destroys this burstiness, spreading events more evenly across time and giving rise to an increased number of temporal paths ending at nodes within short time windows. Note that this result differs from threshold models that do not explicitly require contacts from multiple adopted neighbours. However, there is a clear competing effect arising from correlations between contact trains. Randomly time-shifting the contact events on links was seen to decrease prevalence of adoption. This procedure destroys all correlations between timings of contacts on adjacent links. Because such correlations by design facilitate adoption in our models, the clear decrease in prevalence after randomly offsetting contact trains indicates that such correlations are abundant in the studied networks, as also revealed by direct measurements.

Some features of the dynamics of our threshold models are similar to the behaviour of SI and SIR models on temporal networks~\cite{karsai_smallbutslow,miritello_dynamical_2011}. For example, burstiness hinders the speed and fraction of adopters in both cases, because it increases waiting times along temporal paths. However, there are also noticeable differences. The effect of timing correlations is much more evident for the threshold models, as the threshold dynamics is driven by multiple contacts within a short time window. Further, the sparsity of events and lack of timing correlations may block the adoption process, as seen for the stochastic model applied to the SMS and email networks. For the deterministic model, the adoption in these networks occurs only for very low threshold values, and even then the fraction of adopters is independent of the memory length. Both features are very different from the SI and SIR models, where either adoption always occurs or the final fraction of adopters changes with the model parameters (for SIR, the basic reproduction number). Overall,  threshold models can be seen as an addition to the family of models of contagion in temporal networks. 
\begin{acknowledgments}
  We acknowledge financial support by the Academy of Finland, project
  no:260427.  We thank A.-L. Barab\'asi for the call and SMS data.
\end{acknowledgments}

\bibliography{wilburi.bib}

\end{document}